\documentclass[]{aastex61}
\pdfoutput=1
\usepackage{graphicx}
\usepackage{bm}
\usepackage[version=3]{mhchem}
\usepackage{multirow}

\begin{document}

\title{Studying the Evolution of Warm Dust Encircling BD +20 307 Using SOFIA}

\author{Maggie A. Thompson}
\affiliation{Department of Astronomy and Astrophysics, University of California, Santa Cruz, CA 95064}
\author{Alycia J. Weinberger}
\affiliation{Department of Terrestrial Magnetism, Carnegie Institution of Washington, Washington, D.C., 20008}
\author{Luke Keller}
\affiliation{Department of Physics and Astronomy, Ithaca College, Ithaca, N.Y.}
\author{Jessica A. Arnold}
\affiliation{Department of Terrestrial Magnetism, Carnegie Institution of Washington, Washington, D.C., 20008}
\author{Christopher Stark}
\affiliation{Space Telescope Science Institute, Baltimore, M.D.}

 \email{maapthom@ucsc.edu, weinberger@dtm.ciw.edu}

\begin{abstract}

The small class of known stars with unusually warm, dusty debris disks is a key sample to probe in order to understand cascade models and extreme collisions that likely lead to the final configurations of planetary systems. Because of its extreme dustiness and small radius, the disk of BD +20 307 has a short predicted collision time and is therefore an interesting target in which to look for changes in dust quantity and composition over time. To compare with previous ground and Spitzer Space Telescope data, SOFIA photometry and spectroscopy were obtained.   The system's 8.8-12.5 $\mu$m infrared emission increased by $10 \pm 2 \%$ over nine years between the \textit{SOFIA} and earlier \textit{Spitzer} measurements. In addition to an overall increase in infrared excess, there is a suggestion of a greater increase in flux at shorter wavelengths (less than 10.6 $\mu$m) compared to longer wavelengths (greater than 10.6 $\mu$m). Steady-state collisional cascade models cannot explain the increase in BD +20 307's disk flux over such short timescales. A catastrophic collision between planetary-scale bodies is still the most likely origin for the system's extreme dust; however, the cause for its recent variation requires further investigation.


\end{abstract}

\section{Introduction}

Typical debris disks include leftover planetesimals and are thought to evolve through collisions or evaporation of solid bodies, ranging from small planetesimal to protoplanet/planet-sized \citep{Krivov:2010, Wyatt:2008}.  Just like the Solar System's Kuiper Belt located beyond the orbit of Neptune, most debris disks contain low-temperature dust ($\leq$ 100 K) orbiting far from the host star. However, there exists a small class of known stars with unusually warm, dusty debris disks that serve as a key sample to probe in order to understand cascade models and extreme collisions that likely lead to the final configurations of planetary systems.  Collisional cascade, a process in which larger planetesimals in the disk collide and are continually broken up into smaller objects can explain most debris disks.  In a collisional cascade, small debris disks with warm dust do not last for very long because once the dust has reached a small enough size, removal mechanisms operate quickly, such as radiation pressure that blows the dust out of the system or Poynting-Robertson drag that causes dust particles to fall into the star \citep{Wyatt:2008}.

Compositional changes in warm, dusty disks may be observable on extremely short timescales of years.  As a case in point, observations of TYC 8241 2652 1, a young, Sun-like star with a warm, dusty disk, saw a decrease in the disk's dust emission by a factor of 30 in less than two years, and there is currently no physical model that can explain what would cause such rapid dust depletion \citep{Melis:2012}. Similarly, \cite{Meng:2015} used Spitzer to study five debris disks with unusually high fractional luminosities and found major variations on timescales shorter than a year. 


BD +20 307 can perhaps provide insight into extreme collisional events occurring beyond our Solar System because its warm dust makes it the dustiest star known for its age of $\geq$1 Gyr. The system is a tidally-locked spectroscopic binary composed of nearly identical late-F-type dwarf stars orbiting with a short 3.4-day period, located at 120.0 $\pm$ 0.7 pc from Earth \citep{Weinberger:2008, Fekel:2012, Gaia:2018}.  Two previous instruments obtained infrared spectra for BD +20 307: the first in 2004-2005 using Keck and Gemini at 8--13 $\mu$m \citep{Song:2005}; the second in 2005-2007 using Spitzer at 5--37 $\mu$m \citep{Weinberger:2011}. In addition, long wavelength photometry was collected with Spitzer and in 2011-2012 using Herschel \citep{Vican:2016}. About a decade later in 2015, we obtained new 8--13 $\mu$m infrared spectra using the Stratospheric Observatory for Infrared Astronomy (SOFIA), which is the focus of this paper.  

 For BD +20 307's disk, assuming its $L_{IR}$/$L_{\star}$ = 0.032 represents the surface density of the dust grains, we expect a collision time of only 2.4 years. In addition, since small grains ($\leq$1 $\mu$m) are likely created by collisions, radiation pressure should blow them away on an orbital timescale. Given the old age of the system, collisional cascade cannot explain the large amount of observed dust over the lifetime of BD +20 307 \citep{Wyatt:2007}.  Dust at this flux level cannot last over the system's age, so the dust must be transient. Therefore, another explanation is needed for the cause of the large amount of dust around such a mature system.  In addition, it is important to note that, based on the spectral energy distribution (SED), \cite{Weinberger:2011} and \cite{Vican:2016} conclude that there is no cold dust emitting at far-infrared wavelengths and that the mean temperature of the dust is $\sim$420 K, which suggests an extreme planetary-scale collision within 1 AU to explain BD +20 307's debris disk.

In this paper, we report on the changes in BD +20 307's debris dust by analyzing its spectrum from SOFIA and comparing it to the previous spectra from Keck/Gemini and Spitzer.  In Section 2, we summarize previous observations.  In Section 3, we describe our SOFIA observations of BD +20 307 and the analysis of its new spectra.  We present our results in Section 4, which suggest that we detect significant differences between the SOFIA spectrum and those earlier spectra from Spitzer and the ground-based instruments. In Section 5, we discuss possible mechanisms to explain the observed changes in this extremely warm and dusty debris disk over a decadal timescale. Finally, in Section 6 we summarize our work and propose additional avenues for further analysis of BD +20 307's unusual debris disk system.

\section{Previous Observations}

The first detailed set of observations to study the debris surrounding BD +20 307 were obtained using two ground-based telescopes: the W.M. Keck Observatory and the Gemini-North Telescope, both located on Mauna Kea in Hawaii.  Using the Keck Long Wavelength Spectrometer (LWS) and the Gemini Michelle instrument, \cite{Song:2005} obtained spectra of BD +20 307 taken over the course of three months from August to October of 2004. The resolving powers of the Keck LWS and the Gemini Michelle instrument are R$\approx$150 and $\approx$900, respectively.  \cite{Song:2005} assumed that there was no change in flux over the time period of their observations.  


A year later, the Spitzer Space Telescope gathered the second detailed set of spectroscopy and photometry observations on BD +20 307 \citep{Weinberger:2011}.  On August 20, 2005, Spitzer's Infrared Array Camera (IRAC) gathered short wavelength photometry, and on January 15, 2006, the Infrared Spectrometer (IRS) obtained spectra of BD +20 307. About a year later, on January 21, 2007, the Multiband Imaging Photometer for SIRTF (MIPS) gathered long wavelength photometry of BD +20 307.  The IRS spectral resolution from 8 to 13 $\mu$m was R$\approx$120-600.  \cite{Weinberger:2011} also made the assumption that there were no changes in the flux or spectrum despite the fact that the observations spanned August 2005 to January 2007.

We re-normalized the Spitzer IRS spectrum to the IRAC 5.7 $\mu$m filter instead of to the MIPS 24 $\mu$m filter as in \citet{Weinberger:2011} because the IRAC photometry was gathered closer in time to the IRS spectrum. There are slight differences in the Spitzer IRS spectrum normalized to the IRAC flux compared to the MIPS flux (see Section 2.3 of \citet{Weinberger:2011}). By averaging them over SOFIA's 11.1 $\mu$m filter transmission curve, we found that they differ by 6.8\%.  In the $\sim$ 1.5 years between the Keck/Gemini and Spitzer observations, there is indication that the flux increased slightly over time, particularly at shorter wavelengths. The ratio of the weighted average flux values between the Spitzer and Keck/Gemini spectra for wavelengths less than 10.6 $\mu$m is $1.14\pm0.002$. It is important to note that we have taken the as-published Keck/Gemini data, and there may be additional calibration uncertainties that were accounted for. The uncertainty we have quoted should be considered as a lower bound.


\begin{figure}[ht]
\includegraphics[scale=0.35]{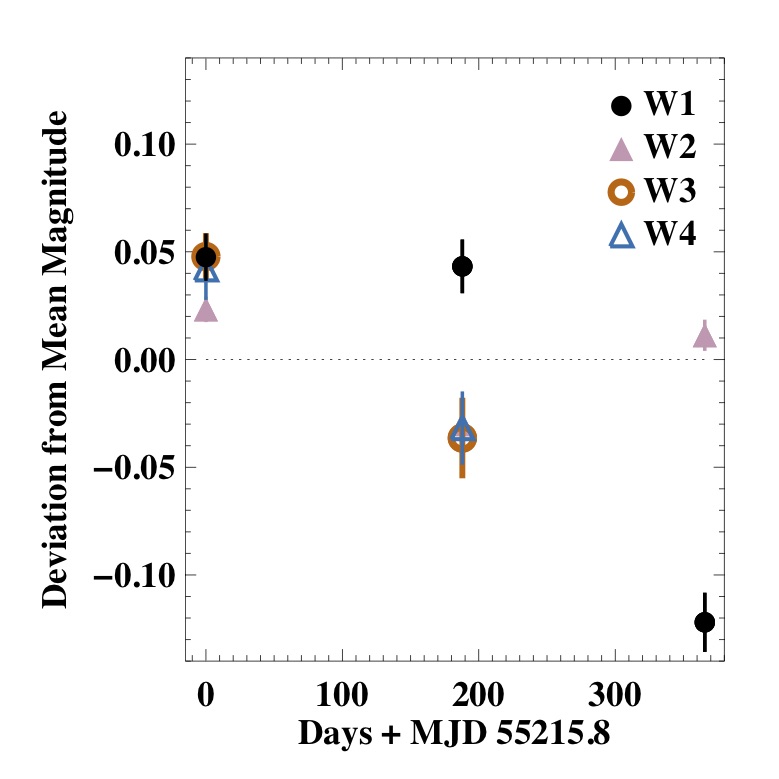}
\centering
\caption{WISE photometry observations of BD +20 307 taken from January 2010 to January 2011. Observations in W3 and W4 only span about 200 days whereas those in W1 and W2 cover the entire year of observations.}
\label{fig:image1}
\end{figure}

In January and July 2010, during the cryogenic mission, WISE measured BD+20 307 at all four of its bands (3.5 - 22 $\mu$m), and in January 2011, it did a scan in the two short wavelength bands. \citet{Meng:2012} noted that BD +20 307's disk may vary by $\sim 7\%$ between the two cryogenic scans taken $\sim$188 days apart but only in the W3 and W4 bands (12 and 22 $\mu$m, respectively). We show the WISE photometry at each epoch in Figure \ref{fig:image1}, and find, in agreement with Meng et al., that all four bands show the same trend, i.e. a small brightening over time, although W1 and W2 are constant within their uncertainties.

In July 2011 and January 2012, the Herschel Space Observatory used its Photodetector Array Camera and Spectrometer (PACS) to measure BD +20 307 at 70 and 100 $\mu$m.  The reported 70 $\mu$m flux density of 47 $\pm$ 3 mJy \citep{Vican:2016} is substantially larger than that reported with MIPS from 2007 of 28.6 $\pm$ 1.9 mJy \citep{Weinberger:2011}.

Most recently, from 2012 to 2013, \cite{Meng:2015} made near-infrared time-series observations of five extreme debris disks including BD +20 307 using Spitzer's IRAC at 3.6 and 4.5 $\mu$m. Although they did not detect a significant trend at 3.6 $\mu$m, at 4.5 $\mu$m they found that BD +20 307's disk flux increased over their period of observations with an average increase rate of $2.5 \pm 0.7$ mJy per year. Such an increase rate coincides within a few percent of the disk flux every year \citep{Meng:2015}. Table \ref{tab:meas} summarizes the main sets of previous infrared observations on BD +20 307's dust.

\begin{table}[ht]
\setlength{\tabcolsep}{0.5pt}
\centering
\begin{tabular}{c c c c c}
\hline
 Observatory & Instrument & Wavelengths ($\mu$m) & Dates & Observation Type \\ 
 \hline
  Keck & LWS & 3.9-24.5 & 08/29/2004  & Photometry \& Spectroscopy \\ 
  Gemini-North & Michelle & 7.7-18.1 & 09-10/2004 & Photometry \& Spectroscopy \\ 
  Spitzer & IRAC & 3.5-7.9 & 08/20/2005 & Photometry  \\ 
  Spitzer & IRS & 5.2-37.2 &  01/15/2006 & Spectroscopy \\ 
  Spitzer & MIPS & 24, 70, 160 & 02/04/2007 & Photometry \\ 
  WISE & & 3.4, 4.6, 12, 22 & 01/19/2010-01/20/2011 & Photometry \\
  Herschel & PACS & 70, 100, 160 & 07/2011, 01/2012 & Photometry \\
  Spitzer & IRAC & 3.6, 4.5 & 2012-2013 & Photometry \\
 \hline
\end{tabular}
\caption{Summary of previous observations of BD +20 307's dust.\label{tab:meas}}
\end{table}


\section{Methods}

\subsection{SOFIA Observations}


For the SOFIA Cycle 2 Program 02\_0050 (PI: Weinberger), we used SOFIA's Faint Object Infrared Camera (FORCAST), a dual-channel mid-infrared camera and spectrograph sensitive to 5-40 $\mu$m. Observations of BD +20 307 were obtained in 2015 flying at an average altitude of 40,000 feet. Tables \ref{tab:spec} and \ref{tab:phot} below summarize the SOFIA observations for gathering spectroscopy and imaging data of BD +20 307. Using the `Nod-Match-Chop' grism observing mode with a chopper throw of 60$''$, data were taken with the `G111' grism that has a nominal resolving power of R=130 for the 4.7" x 191" slit and covers 8.4-13.7 $\mu$m.


\begin{table}[ht]
\setlength{\tabcolsep}{0.5pt}
\centering
\begin{tabular}{c  c  c  c}  
 \hline
  Instrument & Date & Altitude Range (ft) & Integration Time (s) \\ [0.5ex] 
 \hline
 \multirow{3}{5em}{FORCAST `G111'} & 2015 Feb 04 & 39,953-41,014 & 486 \\ [0.5ex]
& 2015 Feb 06 & 39,032-39,031 & 754 \\ [0.5ex]
\hline
 \end{tabular}
 \caption{Spectroscopy observations with SOFIA's FORCAST instrument using the `G111' Grism in the Short Wavelength Channel. \label{tab:spec}}
\end{table}

A total of 55 individual spectra were taken over two nights, 23 on February 4 and 32 on February 6.   We analyzed the products produced by the standard `FORCAST\_REDUX' pipeline version 1.2.0.  The general pipeline process is as follows: load the data and calculate variance; clean bad pixels; correct for applied channel suppression (i.e., droop); correct for image non-linearity; perform background subtraction; remove jailbars; perform spectral extraction using FSpextool algorithm (for spectra only); combine multiple observations (i.e., stacking); calibrate flux. It is important to note that there is a large amount of noise in the data between 9.5-9.9 $\mu$m caused by telluric ozone absorption. The spectrum from each night is available from the SOFIA Data Cycle System as a FITS file that contains the combined spectrum corrected for atmospheric transmission and instrumental response.


\begin{table}[ht]
\setlength{\tabcolsep}{0.5pt}
\centering
 \begin{tabular}{c  c  c  c} 
 \hline
 Instrument & Date & Altitude (ft) & Integration Time (s) \\ [0.5ex] 
 \hline
 \multirow{3}{5em}{FORCAST (Imaging SWC, \\ `F111')} & 2015 Feb 04 & 39955 & 15.61 \\ [0.5ex]
& 2015 Feb 04 & 39953 & 15.61 \\ [0.5ex]
& 2015 Feb 04 & 41013 & 15.73 \\ [0.5ex]
& 2015 Feb 04 & 41010 & 15.73 \\ [0.5ex]
\hline
 \end{tabular}
 \caption{Photometry observations with SOFIA. \label{tab:phot}}
\end{table}

To combine the spectra taken on two different nights, we calculated the mean flux value between 10 and 13 $\mu$m for each of the spectra and normalized the first spectrum to the second spectrum's average flux value in that wavelength region.  We then combined the two spectra via a weighted average in which we weighted each spectral channel by the reciprocal of the squared uncertainty in the flux. In order to determine the variation in the total flux for the two spectra and to ensure that our method of normalization was sound, we averaged the flux over the 11.1 $\mu$m filter (`FOR\_F111') transmission curve to determine the total flux density from each of the spectra and the weighted average spectrum across the 11.1 $\mu$m filter bandpass.  We found that the individual spectra taken two days apart differed from each other by 8.3\%.

In addition to the spectroscopy, target acquisition images were taken in the F111 filter (centered at 11.1 $\mu$m). We used four images out of the 20 taken on February 4.  As with the spectral data, all of the images were processed using `FORCAST\_REDUX.'  For the images, the pipeline includes the same steps mentioned above for spectroscopy but there are several additional steps, including correcting for distortion, merging the images (by shifting and rotating), and finally registering. The pixel scale for the images is 1"=1.3 pixels (or 1 pixel = 0.8").  To average the four images, we determined the centroid of each of the four images and  chose to shift all of the other images to the image having the brightest centroid (i.e., highest signal-to-noise).  We then averaged all four aligned images via a weighted average, once again weighting by the reciprocal of the squared uncertainty of the image (i.e., the reciprocal of the variance associated with each pixel in the image).



\subsection{Photometric Calibration and Normalization of SOFIA Spectrum}

\begin{figure}
\includegraphics[scale=0.55]{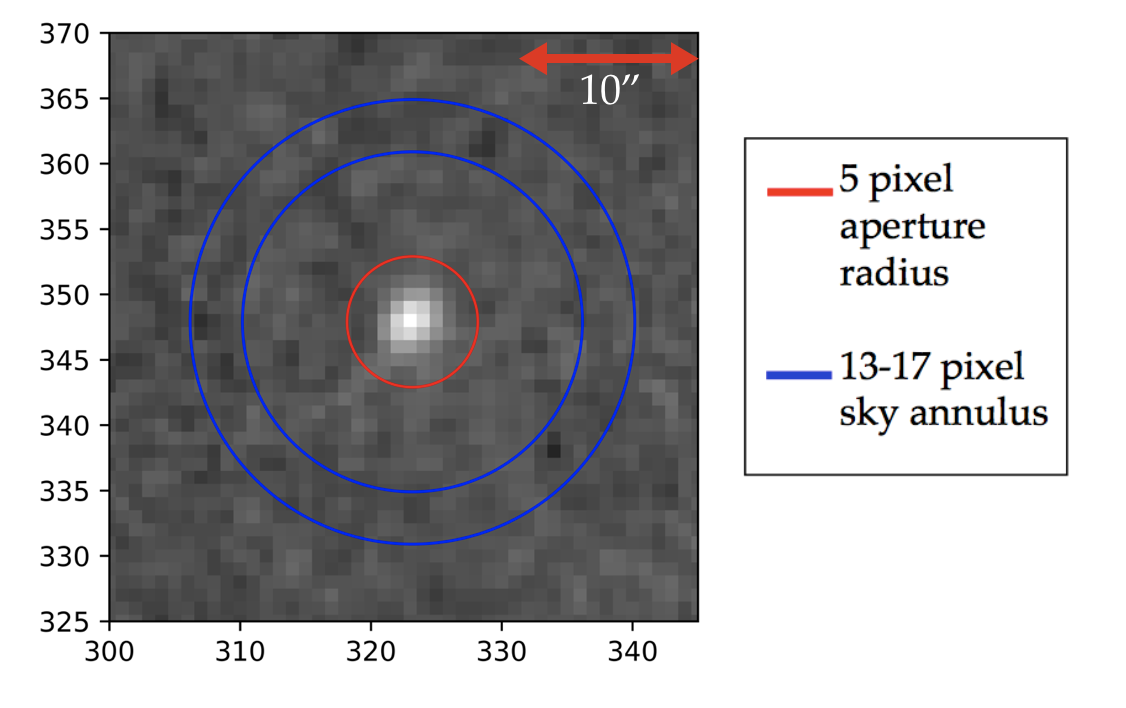}
\centering
\caption{Image of BD +20 307 taken by SOFIA with the red circle representing the 5-pixel aperture radius and the blue circles representing the 13-17 pixel sky annulus we used to perform the aperture photometry.}
\label{fig:image2}
\end{figure}

We photometrically calibrated the SOFIA spectrum by comparing the flux of BD +20 307 from the spectrum to that from the imaging photometry.  We created several curve of growth plots of flux as a function of pixel radius from the center of the source in the image to determine the proper aperture radius and sky annulus that should be used for our aperture photometry calculation. We determined that an aperture radius of 5 pixels and a sky annulus from 13 to 17 pixels was appropriate (Figure \ref{fig:image2}).  Performing aperture photometry, we found that our source had an average flux of 0.41$\pm$0.013 Me/s which is equivalent to 1.07$\pm$0.03 Jy. 

To determine exactly how much flux is missing in our 5-pixel aperture, we did aperture photometry of Alpha Boo, the calibration standard star used for our source.  We compared photometry with the standard aperture radius and sky annulus used in the FORCAST pipeline (i.e., 12 pixel aperture radius, 15-25 pixel sky background region) \citep{FORCASTGIHandbook} and ours (5 pixel radius, 13-17 pixel sky annulus) and determined that our choice returns 83\% of the total flux from a source.  Therefore, to determine the total average flux of our source, we divided 0.41 Me/s by 0.83, which gives the total flux of our source, 0.497 Me/s.  To convert this flux to units of Jansky (Jy), we multiply by the calibration factor determined by the FORCAST pipeline, 0.383 Me/s/Jy, which results in an average flux of 1.30 $\pm$ 0.052 Jy. 

There are four values contributing to the uncertainty of the photometric calibration: first, the photon counting statistics (0.0127 Me/s); second, the fact that we are choosing a specific sky annulus for the aperture photometry (0.01 Me/s); third, the standard deviation of the ratios of flux from the eleven calibration images (0.0118 Me/s); and fourth, the calibration factor uncertainty which is given in the FITS header ($6.5\times10^{-5} Me/s/Jy$). We combined these in quadruture to get a total uncertainty of 0.052 Jy.


Finally, we normalized the SOFIA spectrum to the photometry by computing its flux density over the 11.1 $\mu$m filter transmission curve and then making this equal to our photometric measurement. This resulted in multiplying the SOFIA spectrum by 1.24.  To determine the total uncertainty of BD +20 307's SOFIA flux we combined in quadrature the uncertainty of its photometric flux described above and the uncertainty of its spectral flux (i.e., the uncertainty of each flux value for BD +20 307's SOFIA spectrum).

\subsection{Photosphere Subtraction of the Average SOFIA Spectrum}

To model the stellar flux, we fit a Kurucz model atmosphere of effective temperature of 6000 K and surface gravity (log(g)) of 4.5 to the visible and near-infrared photometry of the star and then extrapolated it to the SOFIA wavelength range. Our photosphere subtraction routine assumes that BD +20 307's stellar spectrum can be treated as that of a single star.  


 
\subsection{Normalization and Photosphere Subtraction of Ground-Based Spectrum}

Before comparing the two earlier epochs of BD +20 307's infrared spectra to its more recent SOFIA spectrum, we first normalized and photosphere subtracted the ground-based spectrum obtained in 2004-2005 using Keck and Gemini. We normalized this spectrum to Keck's Long Wavelength Spectrometer SiC filter whose bandpass is 10.64-12.96 $\mu$m.  From \cite{Song:2005}, BD +20 307's flux at the LWS SiC filter is 843 mJy, we also normalized the spectrum to this flux value. Then, we photosphere subtracted in the same way as for the SOFIA data, above. We also found a new uncertainty for the ground-based spectrum integrated over the SiC filter, taking into account the uncertainty in the photometry reported in \cite{Song:2005}, 0.039 Jy.

\begin{figure}[ht]
\includegraphics[scale=0.055]{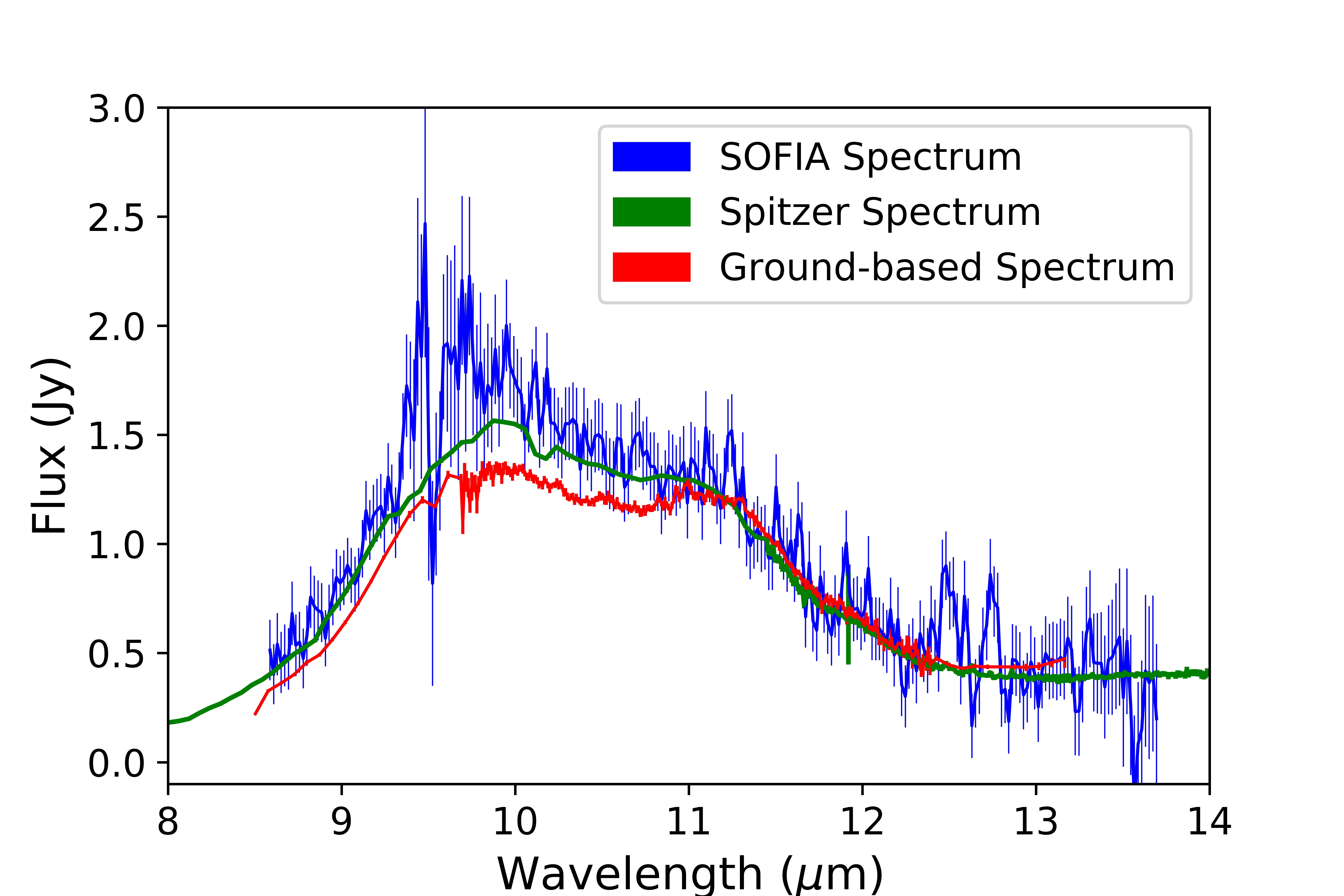}
\centering
\caption{Spectra of BD +20 307 from SOFIA (blue), Spitzer (green), and Keck/Gemini (red) (each have been normalized and photosphere-subtracted).}
\label{fig:image3}
\end{figure}

\section{Results}
 
\begin{figure}[h]
\centering
\includegraphics[scale=0.055]{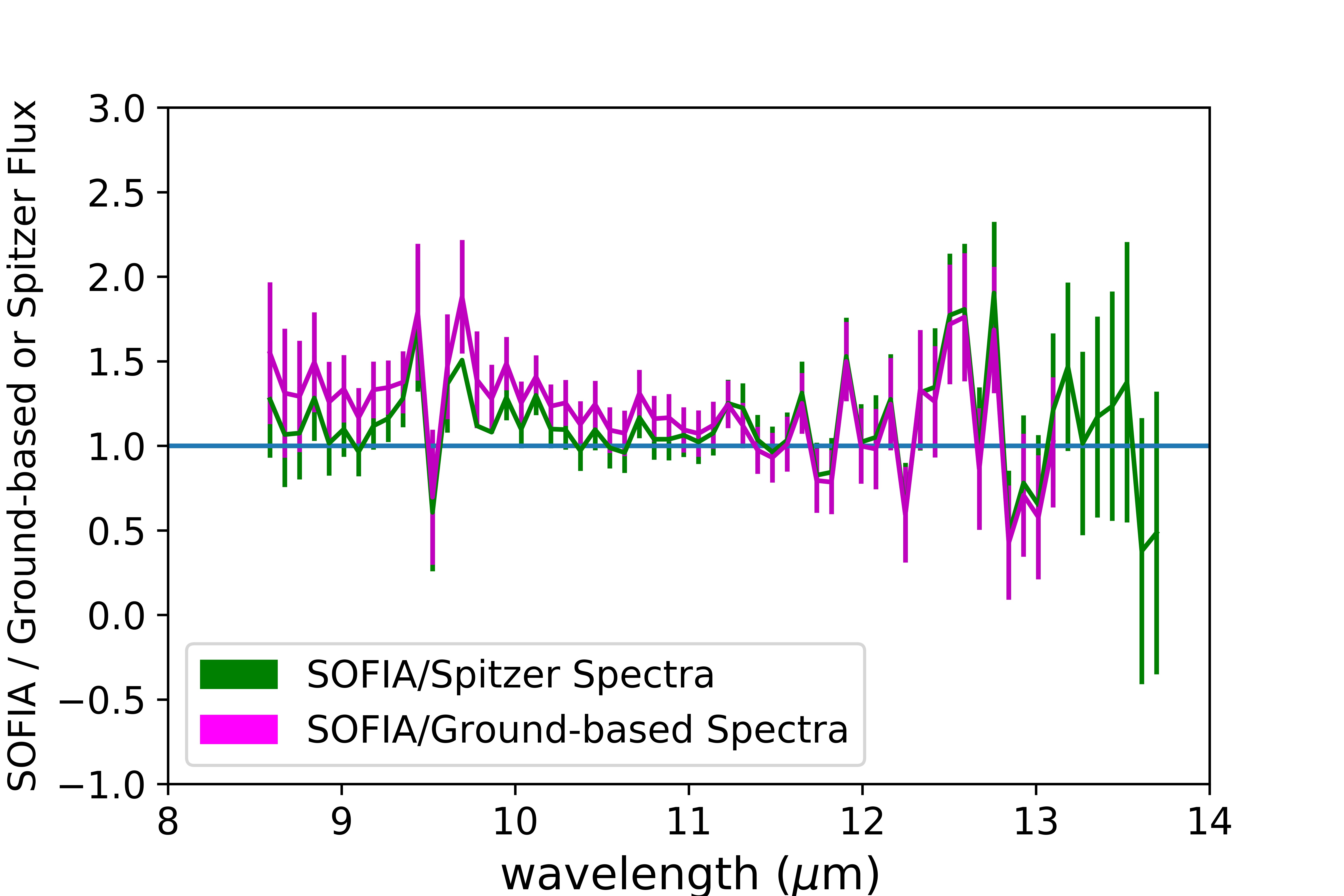}
\caption{Ratio between the SOFIA spectrum and the ground-based and Spitzer spectra. The $Y=1$ line (blue) is what we would expect if there was no difference between SOFIA and the previous spectra.  The average ratio from 8.8 to 12.5 $\mu$m (and excluding the ozone region 9.4-9.9 $\mu$m, which is slightly broader ($\sim$9.4-10.0 $\mu$m) when comparing SOFIA and the ground-based data due to the higher uncertainty in the ground-based ozone calibration) between the SOFIA and Spitzer spectra is 1.08$\pm$0.01 and the average ratio between SOFIA and the ground-based data is 1.15$\pm$0.01, suggesting that the flux has increased over time. It is important to note that the uncertainty in these plots also includes the calibration uncertainty.}
\label{fig:image4}
\end{figure}

We now analyze how the infrared spectrum of BD +20 307 has evolved over the course of $\sim$10 years.  Figure \ref{fig:image3} shows the spectra from earlier observations with Keck/Gemini and Spitzer along with the spectrum from SOFIA. It is important to note that the region between $\sim$9.4-9.9 $\mu$m is where telluric ozone absorption adds noise to the spectrum; therefore, we exclude this region in our analysis. When comparing the SOFIA to the ground-based spectra, we exclude a slightly broader ozone region ($\sim$9.4-10.0 $\mu$m) because the ozone region is less well-calibrated for the ground-based data. Figure \ref{fig:image4} illustrates the ratio between SOFIA's flux and the Keck/Gemini flux (magenta) and SOFIA's flux and the Spitzer flux (green). This plot suggests that the flux from BD +20 307's dust has increased from 2004 to 2015.  In addition to examining the average ratios between the SOFIA and Spitzer spectra we also compare the two by integrating under their spectra.  By comparing the integral under SOFIA's spectrum to that of Spitzer's spectrum (over the same wavelength range), we find that BD +20 307's flux has increased by $10 \pm 2\%$ in the 8 years between their observations.


To determine if there is a change in the \textit{shape} of the spectrum over the last 10 years, we compared the weighted average flux ratio for both SOFIA/Spitzer and SOFIA/Keck-Gemini (green and magenta curves in Figure \ref{fig:image4}, respectively) in two wavelength regimes: below 10.6 $\mu$m and above 10.6 $\mu$m. When calculating these ratios, we only consider flux values with signal/noise ratios greater than 5. We chose 10.6 $\mu$m as the dividing wavelength because 10.6 $\mu$m is the wavelength that most clearly distinguishes between the spectral peaks due to crystalline and amorphous grains (see Fig 2 in \cite{Weinberger:2011}).

\begin{figure}
\centering
\includegraphics[scale=0.055]{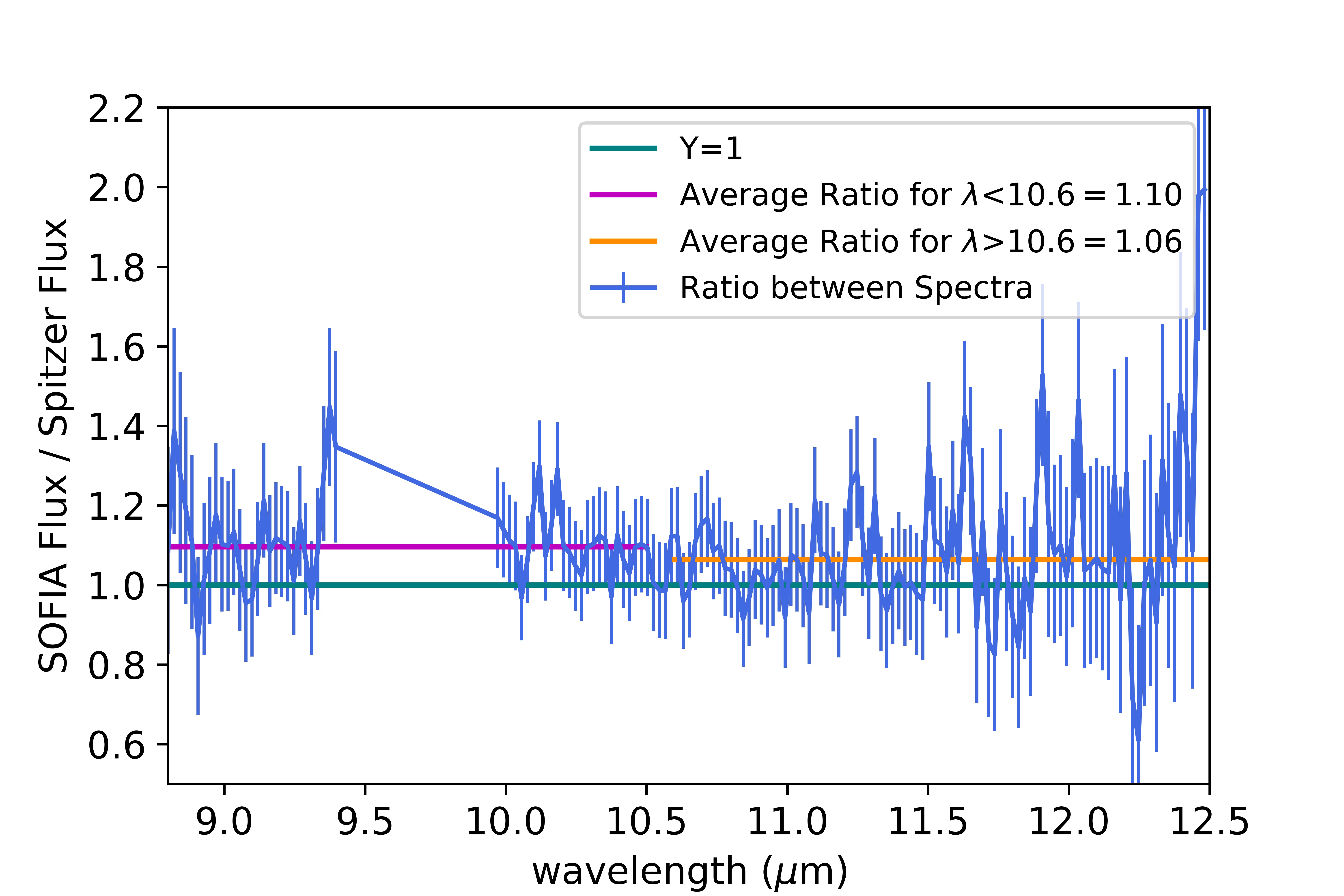}
\caption{Ratio of SOFIA and Spitzer spectra (blue), only including flux values with signal/noise ratios greater than 5. The magenta and orange solid lines represent the weighted average ratio for wavelengths less than 10.6 $\mu$m (and ignoring the ozone region from 9.4 to 9.9 $\mu$m) and wavelengths greater than 10.6 $\mu$m, respectively. The turquoise line (Y=1) is what we would expect if there was no change between the SOFIA and Spitzer spectra. By comparing the ratios of the SOFIA-to-Spitzer spectra in these two regions, it is suggestive of a greater increase in flux over time at shorter wavelengths.}
\label{fig:image5}
\end{figure}

For wavelengths less than 10.6 $\mu$m, we find that the average ratio between the SOFIA and Spitzer spectra is 1.10$\pm$0.02, while for wavelengths greater than 10.6 $\mu$m, the average ratio is 1.06$\pm$0.02. Figure \ref{fig:image5} shows the ratio between the SOFIA and Spitzer flux values (blue curve) and the average ratio values for shorter (magenta line) and longer wavelengths (orange line).  In addition, by comparing the ratio of the integral under the spectrum in the two wavelength ranges (less than and greater than 10.6 $\mu$m) for the SOFIA and Spitzer data: $(\frac{F_{Short}}{F_{Long}})_{SOFIA}=1.77\pm0.52$ vs. $(\frac{F_{Short}}{F_{Long}})_{Spitzer}=1.67\pm0.01$, we find that, while the data are suggestive, the uncertainties in the SOFIA spectrum are too large to allow us to say if there has been a larger increase in flux at shorter wavelengths.

By comparing the integral under SOFIA's spectrum to that of the ground-based spectrum (over the same wavelength range), we determine that BD +20 307's flux has increased by $29 \pm 6\%$ over $\sim$10 years.  While this result suggests an even greater increase in total flux between SOFIA and the Keck/Gemini spectra compared to SOFIA and Spitzer, the ground-based spectra calibration is less reliable compared to the Spitzer calibration. For the ratio between the SOFIA and Keck/Gemini spectra observed 10 years apart, the same general trends are observed, and in fact the differences between the two wavelength regimes is even more pronounced. For wavelengths less than 10.6 $\mu$m, we found the average ratio is 1.25$\pm$0.02, and for wavelengths greater than 10.6 $\mu$m, the average ratio is 1.08$\pm$0.02. The reported uncertainties on these values are dominated by the uncertainties in the SOFIA data, and there are likely additional calibration uncertainties for the Keck/Gemini data that exceed the quoted errorbars. 

We recomputed BD +20 307's disk parameters using the updated stellar luminosity from Gaia \citep{Gaia:2018}. A fit to B and V photometry from Hipparcos, and J, H, and $K_s$ photometry from 2MASS and using the new Gaia parallax of  8.334 $\pm$ 0.046 mas yields a luminosity of 2.9 $L_\odot$. In computing the stellar parameters, we ignore interstellar extinction effects given that BD +20 307 is nearby and not located in the Galactic plane (Galactic latitude = -39$^o$). In addition, according to Schlafly \& Finkbeiner, the extinction in the direction of BD +20 307 is small \citep{Schlafly:2011}.  We assume each star contributed half of this luminosity, has a T$_{eff}$ of 5900~K and [Fe/H]=-0.43 \citep{Zuckerman:2008}, and an age of $\sim$1 Gyr \citep{Weinberger:2008}. These constraints correspond to a mass of $\sim$0.9 - 1 M$_\odot$ on the Dartmouth evolutionary tracks \citep{Dotter:2008}. We take the total mass to be 1.9 M$_\odot$. At the updated luminosity, the dust is located at 1.04 AU.   Otherwise, we use the dust model from \cite{Weinberger:2011} to consider how the disk may have changed over time. That model assumes the dust to be at a single distance from the stars and made up of four compositions: amorphous olivine, crystalline olivine, amorphous pyroxene and large blackbody grains (see Section 3.2 in \cite{Weinberger:2011} for more details).

\section{Discussion}

Our analysis reveals that between $\sim$8.8-12.5 $\mu$m BD +20 307's disk dust flux increased by $\sim10\pm2\%$ over the eight years between the Spitzer and SOFIA measurements and $\sim29\pm6\%$ over the $\sim$10 years between the Keck/Gemini and SOFIA observations. If we assume the disk is optically thin and all of the dust grains have the same size (0.5 $\mu$m) and opacity, then to get a 10\% increase in luminosity between the SOFIA and Spitzer observations by just increasing the number of dust grains, we would need to introduce $\sim9*10^{32}$ more grains. If we imagine combining all of the dust grains into one spherical object, then the additional dust grains required to increase the luminosity by 10\% would make up a 48 km radius body compared to the $\sim$110 km radius object that would contain enough mass for the grains required to get the $L_{IR}/L_{\star} = 0.032$  \citep{Weinberger:2011}. It is important to acknowledge that this number of additional dust grains is a lower limit because we assume an optically thin disk. If, however, the disk were optically thick as some slightly younger debris disks can be, we would need to add even more dust grains to explain the change in flux.

\begin{figure*}
\includegraphics[trim={2cm 0 2cm 0},clip, scale=0.65, angle=90]{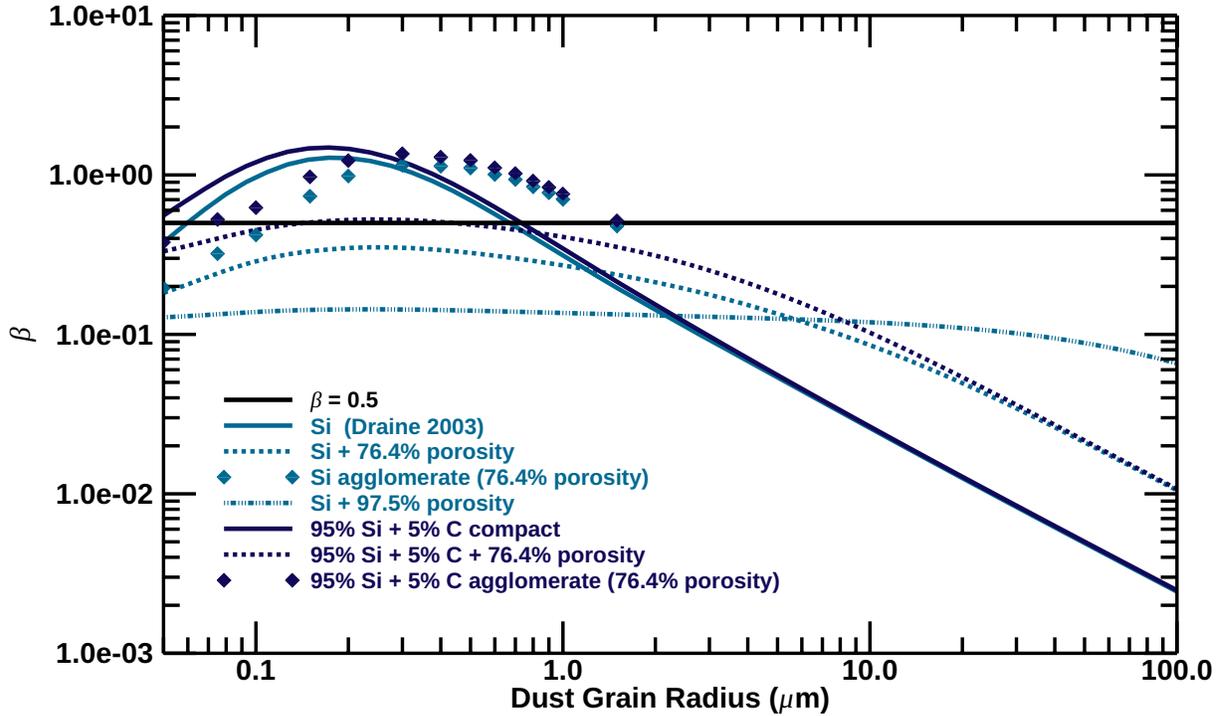}
\centering
\caption{$\beta$-ratio vs. dust grain radius for pure astro-silicate (teal) and an astro-silicate amorphous carbon mixture (indigo). For each of these two compositions, $\beta$-ratios for compact Mie spheres (solid lines) are compared with 76.4\% porous Mie spheres modified using the Bruggeman mixing rule (dashed lines) and 76.4\% porous irregular agglomerated debris particles. The method for generating agglomerated debris particles is described in \cite{Zubko:2005}. This porosity was chosen based on the average for agglomerated debris particles as calculated by \cite{Zubko:2015}. For pure astro-silicate, an extremely porous Mie sphere example (97.5\%) is also shown (dash-dotted line). Scattering properties for the irregular agglomerated debris particles are averaged over at least 500 randomly generated and oriented particles, with particles added as necessary until the variation is less than 1\%.}
\label{fig:beta}
\end{figure*}

Understanding how BD +20 307's dust flux can be increasing on such short timescales requires that we first discuss why this system is so unusual and does not conform to typical debris disk evolution models. Given the mature age ($\geq1$ Gyr) of BD +20 307, it is extremely unusual for the system to have such copious amounts of warm dust ($L_{IR}$/$L_\star$ = 0.032) within $\sim$1 AU. The most likely explanation for the origin of extremely dusty debris disks is an extreme collision between planetary-scale bodies. After such a catastrophic collision, the general assumption for dust evolution is a steady-state collisional cascade, a process by which planetesimals collide and break into smaller objects which in turn collide and continuously grind down to smaller and smaller fragments (e.g., \cite{Thebault:2003}, \cite{DominikDecin:2003}, \cite{Wyatt:2007}). Once objects break apart to small enough sizes, they will be removed either by radiation pressure pushing the particles out of the system or Poynting-Robertson drag pulling them into the star \citep{Wyatt:2008}. There is an important caveat to this which is that in some cases, depending on the ratio of the radiation pressure to the stellar gravitational force as a function of grain size, very small grains will not get removed by either process; and in other cases, grains will never get blown out by radiation pressure (see Figure \ref{fig:beta} and further discussion below).  Nevertheless, the collisional timescale for a system as dusty as BD +20 307 is still too short (several thousand years) to sustain so much warm debris over the age of the binary stars \citep{Wyatt:2007}. Furthermore, it seems unlikely that we just happened to catch the initial start of the collisional cascade when we began observing the debris disk system only thirty years ago. Even if we did serendipitously start observing BD +20 307 at the beginning of the cascade, a purely steady-state collisional cascade model cannot explain why we see variations in the dust flux, and especially an \textit{increase} in the infrared excess, over such short yearly timescales.  Therefore, in order to explain the variations we observe, we must relax some of the assumptions made in steady-state collisional cascade.

Under the model for steady-state cascade evolution, the size distribution of dust grains follows a power-law (usually $f(a) \propto a^{\gamma}$, where $\gamma$ = -3.5) 
\citep{DominikDecin:2003}. This power-law assumption is usually assumed to work down to an abrupt small-size cutoff in the distribution caused by the characteristic blowout size for small dust grains (as shown in Figure \ref{fig:beta}) which introduces waves into the dust grain size distribution (\cite{CampoBagatin:1994}, \cite{Thebault:2003}).  As Figure \ref{fig:beta} illustrates, for a given dust grain composition, there is a characteristic blowout size (denoted by where a given curve intersects the $\beta$=0.5 line) below which grains will be ejected from the system determined by $\beta$, the ratio between the radiation pressure pushing grains out and gravity from the host stars pulling them in.  We consider grains made of either pure astro-silicate or an astro-silicate amorphous carbon mixture, and for each of these grain compositions we consider three types of grains: compact spheres, porous spheres and agglomerated debris particles (see \citet{Arnold:2019}). In the case of this binary system, very small grains can persist in the disk, and these can be very effective at breaking up larger grains and launching avalanches that may happen stochastically over time (\citet{Krivov:2010}, \citet{Grigorieva:2007}). The persistence of small grains also helps explain why the silicate feature is so prominent even though the canonical blowout size is $>$1 micron (see \citet{Weinberger:2011}). One could also imagine adding even more complexity to collisional cascade models, in which, for example, there are planetesimals of different strengths because of different compositions or thermal histories rather than just size.  In addition, planetesimal strength can vary depending on their size distribution \citep{ObrienGreenberg:2003}. Relaxing assumptions like these in the steady-state cascade model may help explain why we see some variation in the dust flux over short timescales.

\begin{figure*}
\includegraphics[scale=0.7]{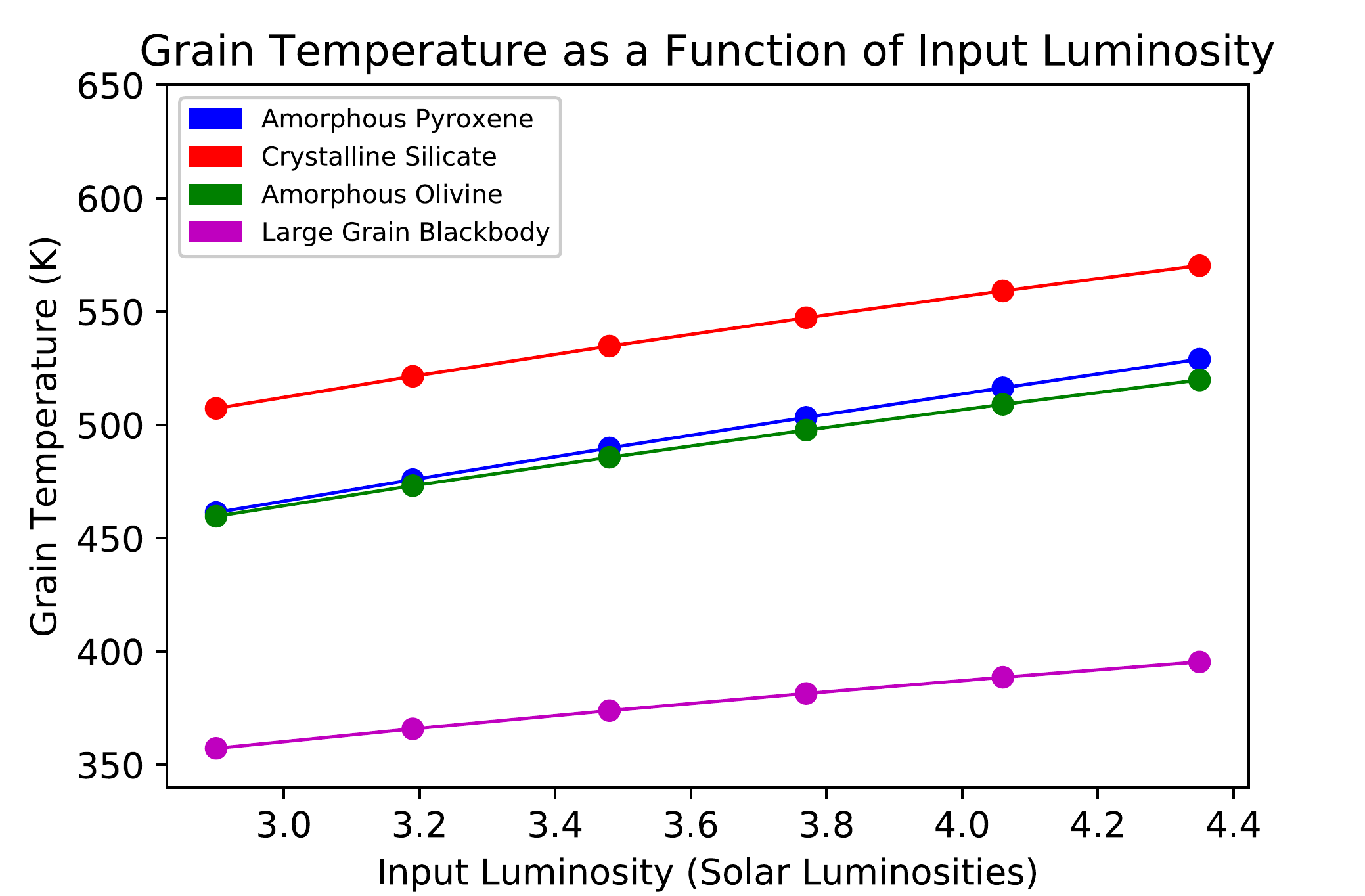}
\centering
\caption{Grain temperature as a function of stellar luminosity for four grain compositions. For each composition, the grain temperature increases as input luminosity from the star increases.}
\label{fig:temp}
\end{figure*}

There are two possible causes for a flux increase: an increase in temperature of the dust (since $L \propto T^{4}$) or an increase in the surface area of the disk that is visible to the observer.  In our model of \cite{Weinberger:2011}, the dust grains sit in a ring and have temperatures that vary according to their absorptivities, from 357~K for blackbody grains to 507~K for crystalline silicates.  There are two ways the dust temperature could increase: increasing the stellar flux or moving the dust closer to the binary stars. The dust temperature would increase if the stellar flux at wavelengths where the dust absorbs strongly (UV-visible) increased. To increase the flux by 10\% takes a temperature increase of $\sim$3.5\% or a luminosity increase of 12\%. Figure \ref{fig:temp} shows the impact of increasing the stellar luminosity on the temperature of varying compositions of dust grains. We note that this would also produce a small change in the radiation pressure blowout size ($\beta \propto L_{*}$; see below and Figure \ref{fig:beta}). Increasing the stellar luminosity by 12\% would increase the grain temperatures to 475, 478 and 524 K (from 459, 461 and 507 K) for amorphous olivine, amorphous pyroxene and crystalline olivine, respectively (Figure \ref{fig:temp}, \cite{Weinberger:2011}). These temperatures are well below the melting temperatures, and thus would not change the composition or grain size distribution of the dust. 

Although typical single F-type stars are not very variable at old ages, F-type binary stars can be variable, despite the fact that their activity is believed to decrease with age \citep{Donahue:1998}. For instance, HD 212280 is a chromospherically active spectroscopic binary of similar age and spectral types as BD +20 307 and was discovered to be variable through photometric monitoring, with several starspots being the assumed cause for the variability \citep{Fekel:1993}.  BD +20 307 is not known for certain to be variable at visible wavelengths; however, once \textit{Gaia} gathers all its data, it will be a good source to check for stellar activity. \cite{Zuckerman:2008} measured BD +20 307's X-ray flux in the 0.5-2.0 keV band to be $1.1\times10^{-13} erg/cm^2/s$, corresponding to a luminosity of $4.7\times10^{-5}$ $L_{\odot}$ and a fractional X-ray luminosity ($\frac{L_X}{L_*}$) of $1.6\times10^{-5}$, which alone cannot cause an $\sim$10\% increase in the stellar luminosity.  One possible way to get an increase in flux is if the disk is misaligned with respect to the orbital plane of the stars and is precessing on short timescales ($\sim$ years). Assuming hotspots on the stars are generating the X-ray emission, then in the case of a precessing disk, the disk will see the spots periodically and heat up during those times when it is exposed to the spots. Unfortunately, the inclination of BD + 20 307's debris disk is not known, but \cite{Zuckerman:2008} suggest that mid-infrared interferometric measurements might be able to prove whether or not the disk is inclined with respect to the orbital plane of the stars.

The second way to increase the temperature of the dust grains is by moving them closer to the binary stars ($\propto \sqrt d$).  Most debris disks are thought to be in a radiation pressure dominated regime, where collisions grind dust to smaller sizes until the grains reach the blowout size and are ejected on an orbital timescale (e.g., \cite{DominikDecin:2003}, \cite{Wyatt:2007}, \cite{Wyatt:2008}). This mechanism can explain the decline in the incidence of debris disks over time. However, the changes in BD +20 307 are in the direction of \textit{increasing} the disk flux. At the new combined stellar luminosity of 2.9 L$_\odot$, the ring is at 1.04 AU. For the dust flux to increase 10\% requires moving the dust {\it in} $\sim$5\% to 0.99 AU, which increases the temperatures of the dust grains by $\sim$15 K. The change in temperatures is similar amongst the different grain compositions, so that the shape of the spectrum is not much changed by such an increase in temperature. Comparing the ratio of the integral under the spectrum in the short and long wavelength regimes (less than and greater than 10.6 $\mu$m, respectively) for model spectra with the dust at 0.98 AU and 1.04 AU, we find there would be a $\sim$1.4\% increase at the short wavelengths relative to the long wavelengths. 


The Poynting-Robertson drag timescale is given by 400$d^2/\beta$ yr where d is the distance of the grain from the star \citep{Burns:1979}.  As shown in Figure \ref{fig:beta}, we calculated $\beta$ for compact spheres (Mie theory), spheres with porosity (Mie theory plus a Bruggeman mixing rule), and agglomerated debris particles \citep{Zubko:2005} assuming a combined stellar mass of 1.9 $M_\odot$ and luminosity of 2.9 L$_\odot$ \citep{Arnold:2019}.  For grains just larger than the blowout level of $\beta$=0.5, i.e., $\sim$2$\mu$m,  at $\sim$1 AU, Poynting-Robertson drag can bring a grain all the way to the star in 800 yr, or a few tenths of an AU in $\sim$100 yr. This is not fast enough to account for the change we see. In addition, the short collisional timescale would favor radiation pressure blowout.

Stellar wind drag can operate efficiently on small grains, even below the nominal blowout size because small grains have a greater coupling efficiency to the wind. The wind drag timescale can be considered a multiple of the Poynting-Robertson timescale, i.e.  
$t_{sw} = t_{PR} \frac{Q_{PR}}{Q_{sw}}\frac{L_*}{\dot{M}c^2}$ \citep{Plavchan:2005}. Here $Q_{PR}$ and $Q_{sw}$ are the efficiencies of the two processes that depend on grain properties and whose ratio can be assumed to be $\sim$1, and $\dot{M}$ is the wind mass loss rate in Solar masses per year.
For the Sun-like stars, X-ray activity and winds are likely correlated \citep{Feigelson:1982}, and flaring X-ray activity is correlated with stellar coronal mass ejections \citep{Aarnio:2012}. The enhanced X-ray activity of the tidally locked spectroscopic binary in this system might also produce a significant stellar wind. A wind rate that is 10$\times$ Solar would reduce the timescale to drag in dust a few tenths of an AU by an order of magnitude, to $\sim$10 yr, a scale comparable to the time between our observations.


To account for the increased disk flux, it is also possible that the disk's surface area visible to the observer increased either by the production of more dust grains or a change in the optical depth which exposes more dust to the stars and the observer. The total number of grains in the disk would have to increase by about the same fraction as the increase in flux.  This is similar to the ``avalanche" type events considered by \citet{Grigorieva:2007} in which the breakup of a large planetesimal happens in the background of a pre-existing disk of small particles, thus triggering yet more collisions that significantly increase the amount of small dust grains. The ``parent'' bodies that break up in sequential avalanches in this scenario could be fragments of a larger differentiated body that was involved in the original giant collision. The largest body involved in the collision would have to have a timescale for destruction by smaller grains of $\sim$8 years, i.e. the difference in the observations, along with removal of the previously produced small grains on the same timescale. The amount of dust, $\sim$10$^{21}$g, is comparable to that used in the calculations of \citet{Grigorieva:2007}, but the disk is closer to the star and smaller radially, which helps to cause a faster collisional cascade, although exact models have not been done for this regime. However, more recent work on collisional avalanches by Thebault \& Kral find that the photometric excess due to an avalanche, even in the case of extreme debris disks like BD +20 307, is less than 10\%. They claim that while a larger initial planetesimal mass could result in greater photometric excesses, it is less likely for a larger mass object to get broken apart \citep{Thebault:2018}. That being said, determining the frequency at which planetesimals break apart depends strongly on the specific debris disk system such as the size distribution of large objects and their dynamical excitation \citep{Thebault:2018}. Therefore, it would be beneficial to apply Thebault \& Kral's updated collisional avalanche models to a BD +20 307-like system under various assumptions of planetesimal size distribution.


We also found that the flux at shorter wavelengths has increased more than the flux at longer wavelengths over the last 10 years. This possible increase in dust flux, particularly pronounced at wavelengths less than 10.6 $\mu$m, suggests that the dust's silicate composition has possibly evolved during this short timescale. If future data can definitively find a change in the \textit{shape} of the spectrum, there are possible explanations for greater increase in flux at shorter wavelengths. The process of converting from the crystalline to amorphous phase via irradiation has been previously studied in detail (e.g, \cite{Brucato:2004}, \cite{Christoffersen:2011}), however the conversion from crystalline to amorphous via \textit{collisions} has not been as well studied (see \cite{Gleason:2017} as an example of recent experimental work on this conversion). A catastrophic collision is known to subject bodies to extremely high temperatures and pressures. If the temperatures are high enough that the grains reach their melting temperatures and the objects cool at a fast enough rate, which seems plausible in a gas-less disk like BD +20 307's, then they can be quenched into a glass (e.g., \cite{Birnie:1986} studied the cooling rates required to form the silicate glasses on the Moon). The cooling rate must be faster than the tens of minutes it takes for grains to anneal from amorphous to crystalline as determined by \cite{Hallenbeck:2000}. This seems likely given the very short amount of time (nanoseconds) \cite{Gleason:2017} predict that it takes for crystallized silica to convert to an amorphous phase as a result of shock loading induced by an extreme collision. Therefore, if such a collision can indeed induce the conversion from objects' crystalline to amorphous phases, this could provide a possible explanation for the possible greater change in flux at shorter wavelengths.

\section{Conclusions}

We have detected significant variation in BD +20 307's dust flux over the course of $\sim$10 years, from the earlier sets of observations (Keck/Gemini in 2004-2005 and Spitzer in 2006-2007) to our recent SOFIA observations in 2015.  BD +20 307's dust flux has increased by 10\% between $\sim$8.8-12.5 $\mu$m. In addition, while there is indication that the shape of the spectrum has changed, with the dust flux potentially increasing more at shorter wavelengths (less than 10.6 $\mu$m) compared to longer wavelengths (greater than 10.6 $\mu$m), we cannot conclude this definitively due to the high uncertainties in the SOFIA spectra. New SOFIA observations covering a wider wavelength range out to 20 $\mu$m along with fitting composition models of the dust grains to the wider wavelength-range of data (as in \cite{Weinberger:2011}) will allow us to draw more concrete conclusions regarding the most likely cause of this dust flux increase over such short timescales. Nevertheless, it is clear that steady-state collisional cascade alone cannot explain these variations, and particularly the \textit{increase} in dust flux, over such short timescales. Therefore, we must relax some of the assumptions of the steady-state model, such as the size distribution of dust grains following a power-law all the way down to a single minimum grain size, or that all planetesimals have the same strength. We investigated several mechanisms that could cause the observed changes in the disk flux, including making the dust grains hotter, either through an increase in stellar luminosity or moving the dust grains closer to the stars, or increasing the number of dust grains in the system. If the origin of the copious amount of warm dust orbiting BD +20 307 is an extreme collision between planetary-sized bodies, then this system may help unlock clues into planetary systems around binary stars, along with providing a glimpse into catastrophic collisions occurring late in a planetary system's history.  Understanding BD +20 307 and other systems like it with extremely dusty debris disks could advance our knowledge of catastrophic collisions, the effects of binary stars on debris disks and the evolution of planetary systems.

This publication is based on observations made with the NASA/DLR Stratospheric Observatory for Infrared Astronomy (SOFIA). SOFIA is jointly operated by the Universities Space Research Association, Inc. (USRA), under NASA contract NNA17BF53C, and the Deutsches SOFIA Institut (DSI) under DLR contract 50 OK 0901 to the University of Stuttgart. Financial support for this work was provided by NASA through award \#02-0050 issued by USRA.  This publication makes use of data products from the Wide-field Infrared Survey Explorer, which is a joint project of the University of California, Los Angeles, and the Jet Propulsion Laboratory/California Institute of Technology, funded by the National Aeronautics and Space Administration. This research has made use of the NASA/ IPAC Infrared Science Archive, which is operated by the Jet Propulsion Laboratory, California Institute of Technology, under contract with the National Aeronautics and Space Administration.  This work has made use of data from the European Space Agency (ESA) mission {\it Gaia} (\url{https://www.cosmos.esa.int/gaia}), processed by the {\it Gaia} Data Processing and Analysis Consortium (DPAC, \url{https://www.cosmos.esa.int/web/gaia/dpac/consortium}). Funding for the DPAC has been provided by national institutions, in particular the institutions participating in the {\it Gaia} Multilateral Agreement.  We would like to thank Inseok Song for useful preparatory discussions.

\facility{SOFIA, Spitzer, Keck, WISE, Herschel, IRSA}

\bibliography{main}

\begin{thebibliography}{}
\expandafter\ifx\csname natexlab\endcsname\relax\def\natexlab#1{#1}\fi

\bibitem[{Aarnio {et~al.}(2012)Aarnio, Matt, \& Stassun}]{Aarnio:2012}
Aarnio, A.~N., Matt, S.~P., \& Stassun, K.~G. 2012, The Astrophysical Journal,
  760, 11

\bibitem[{Arnold {et~al.}(2019)Arnold, Weinberger, Videen, \&
  Zubko}]{Arnold:2019}
Arnold, J.~A., Weinberger, A.~J., Videen, G., \& Zubko, E.~S. 2019, in prep

\bibitem[{Bagatin {et~al.}(1994)Bagatin, Cellino, Davis, Farinella, \&
  Paolicchi}]{CampoBagatin:1994}
Bagatin, A.~C., Cellino, A., Davis, D., Farinella, P., \& Paolicchi, P. 1994,
  Planetary and Space Science, 42, 13

\bibitem[{Birnie \& Dyar(1986)}]{Birnie:1986}
Birnie, D.~P., \& Dyar, M.~D. 1986, Journal of Geophysical Research, 91, 5

\bibitem[{Brucato {et~al.}(2004)Brucato, Strazzulla, Baratta, \&
  Colangeli}]{Brucato:2004}
Brucato, J., Strazzulla, G., Baratta, G., \& Colangeli, L. 2004, Astronomy \&
  Astrophysics, 413, 6

\bibitem[{Burns {et~al.}(1979)Burns, Lamy, \& Soter}]{Burns:1979}
Burns, J.~A., Lamy, P.~L., \& Soter, S. 1979, Icarus, 40, 48

\bibitem[{Christofferson \& Keller(2011)}]{Christoffersen:2011}
Christofferson, R., \& Keller, L.~P. 2011, Meteoritics \& Planetary Science,
  46, 19

\bibitem[{Dominik \& Decin(2003)}]{DominikDecin:2003}
Dominik, C., \& Decin, G. 2003, The Astrophysical Journal, 598, 9

\bibitem[{Donahue(1998)}]{Donahue:1998}
Donahue, R.~A. 1998, Cool Stars, Stellar Systems and the Sun, ASP Conference
  Series, 154, 8

\bibitem[{Dotter {et~al.}(2008)Dotter, Chaboyer, Jevremović, Kostov, Baron, \&
  Ferguson}]{Dotter:2008}
Dotter, A., Chaboyer, B., Jevremović, D., {et~al.} 2008, The Astrophysical
  Journal Supplement Series, 178, 12

\bibitem[{Feigelson(1982)}]{Feigelson:1982}
Feigelson, E.~D. 1982, Icarus, 51, 8

\bibitem[{Fekel {et~al.}(2012)Fekel, Cordero, Galicher, Zuckerman, Melis, \&
  Weinberger}]{Fekel:2012}
Fekel, F., Cordero, M., Galicher, R., {et~al.} 2012, The Astrophysical Journal,
  749, 11

\bibitem[{Fekel {et~al.}(1993)Fekel, Browning, Henry, Morton, \&
  Hall}]{Fekel:1993}
Fekel, F.~C., Browning, J.~C., Henry, G.~W., Morton, M.~D., \& Hall, D.~S.
  1993, The Astronomical Journal, 105, 11

\bibitem[{{FORCAST Science Instrument Team}(2016)}]{FORCASTGIHandbook}
{FORCAST Science Instrument Team}. 2016, Guest Investigator Handbook for
  FORCAST Data Products, revision b edn., SOFIA

\bibitem[{{Gaia Collaboration} {et~al.}(2018){Gaia Collaboration}, {Brown},
  {Vallenari}, {Prusti}, {de Bruijne}, {Babusiaux}, \&
  {Bailer-Jones}}]{Gaia:2018}
{Gaia Collaboration}, {Brown}, A.~G.~A., {Vallenari}, A., {et~al.} 2018, ArXiv
  e-prints, arXiv:1804.09365

\bibitem[{Gleason {et~al.}(2017)Gleason, Bolme, Lee, Nagler, Galtier, Kraus,
  Sandberg, Yang, Langenhorst, \& Mao}]{Gleason:2017}
Gleason, A., Bolme, C., Lee, H., {et~al.} 2017, Nature Communications, 8, 6

\bibitem[{Grigorieva {et~al.}(2007)Grigorieva, Artymowicz, \&
  Th{\'{e}}bault}]{Grigorieva:2007}
Grigorieva, A., Artymowicz, P., \& Th{\'{e}}bault, P. 2007, Astronomy and
  Astrophysics, 461, 537

\bibitem[{Hallenbeck {et~al.}(2000)Hallenbeck, III, \&
  Nelson}]{Hallenbeck:2000}
Hallenbeck, S.~L., III, J. A.~N., \& Nelson, R.~N. 2000, The Astrophysical
  Journal, 535, 9

\bibitem[{Krivov(2010)}]{Krivov:2010}
Krivov, A.~V. 2010, Research in Astronomy and Astrophysics, 10, 383

\bibitem[{Melis {et~al.}(2012)Melis, Zuckerman, Rhee, Song, Murphy, \&
  Bessell}]{Melis:2012}
Melis, C., Zuckerman, B., Rhee, J.~H., {et~al.} 2012, Nature, 1207.1162

\bibitem[{Meng {et~al.}(2012)Meng, Rieke, Su, Ivanov, Vanzi, \&
  Rujopakarn}]{Meng:2012}
Meng, H. Y.~A., Rieke, G.~H., Su, K. Y.~L., {et~al.} 2012, The Astrophysical
  Journal Letters, 751, 5

\bibitem[{Meng {et~al.}(2015)Meng, Su, Rieke, Rujopakarn, Myers, Cook, Erdelyi,
  Maloney, McMath, Persha, Poshyachinda, \& Reichart}]{Meng:2015}
Meng, H. Y.~A., Su, K. Y.~L., Rieke, G.~H., {et~al.} 2015, The Astrophysical
  Journal, 805, 1503.05610

\bibitem[{O`Brien \& Greenberg(2003)}]{ObrienGreenberg:2003}
O`Brien, D.~P., \& Greenberg, R. 2003, Icarus, 164, 11

\bibitem[{Plavchan {et~al.}(2005)Plavchan, Jura, \& Lipscy}]{Plavchan:2005}
Plavchan, P., Jura, M., \& Lipscy, S.~J. 2005, The Astrophysical Journal, 631,
  8

\bibitem[{Schlafly \& Finkbeiner(2011)}]{Schlafly:2011}
Schlafly, E.~F., \& Finkbeiner, D.~P. 2011, The Astrophysical Journal, 737, 13

\bibitem[{Song {et~al.}(2005)Song, Zuckerman, Weinberger, \&
  Becklin}]{Song:2005}
Song, I., Zuckerman, B., Weinberger, A.~J., \& Becklin, E.~E. 2005, Nature,
  436, 363

\bibitem[{Thebault {et~al.}(2003)Thebault, Augereau, \& Beust}]{Thebault:2003}
Thebault, P., Augereau, J., \& Beust, H. 2003, Astronomy \& Astrophysics, 408,
  13

\bibitem[{Thebault \& Kral(2018)}]{Thebault:2018}
Thebault, P., \& Kral, Q. 2018, Astronomy \& Astrophysics, 609, 13

\bibitem[{Vican {et~al.}(2016)Vican, Schneider, Bryden, Melis, Zuckerman, Rhee,
  \& Song}]{Vican:2016}
Vican, L., Schneider, A., Bryden, G., {et~al.} 2016, The Astrophysical Journal,
  833

\bibitem[{Weinberger(2008)}]{Weinberger:2008}
Weinberger, A.~J. 2008, The Astrophysical Journal Letters, 679, L41

\bibitem[{Weinberger {et~al.}(2011)Weinberger, Becklin, Song, \&
  Zuckerman}]{Weinberger:2011}
Weinberger, A.~J., Becklin, E.~E., Song, I., \& Zuckerman, B. 2011, The
  Astrophysical Journal, 726, 72

\bibitem[{Wyatt(2008)}]{Wyatt:2008}
Wyatt, M.~C. 2008, Annual Review of Astronomy \& Astrophysics, 46, 339

\bibitem[{Wyatt {et~al.}(2007)Wyatt, Smith, Greaves, Beichman, Bryden, \&
  Lisse}]{Wyatt:2007}
Wyatt, M.~C., Smith, R., Greaves, J.~S., {et~al.} 2007, The Astrophysical
  Journal, 658, 569

\bibitem[{Zubko {et~al.}(2005)Zubko, Petrov, Shkuratov, \& Videen}]{Zubko:2005}
Zubko, E., Petrov, D., Shkuratov, Y., \& Videen, G. 2005, Applied Optics IP,
  44, 6

\bibitem[{Zubko {et~al.}(2015)Zubko, Videen, Hines, Shkuratov, Kaydash,
  Muinonen, Knight, Sitko, Lisse, Mutchler, Wooden, Li, \&
  Kobayashi}]{Zubko:2015}
Zubko, E., Videen, G., Hines, D.~C., {et~al.} 2015, Planetary and Space
  Science, 118, 25

\bibitem[{Zuckerman {et~al.}(2008)Zuckerman, Fekel, Williamson, Henry, \&
  Muno}]{Zuckerman:2008}
Zuckerman, B., Fekel, F., Williamson, M., Henry, G., \& Muno, M. 2008, The
  Astrophysical Journal, 688, 7

\end{thebibliography}

\end{document}